\documentclass[final,5p,times,twocolumn]{elsarticle}

\usepackage{amssymb}
\usepackage{lineno}
\usepackage{color}
\usepackage{amsmath}

\journal{Physics Letters B}

\begin{document}

\begin{frontmatter}



\title{Direct dark matter search by annual modulation in XMASS-I}


\author[ICRR,IPMU]{K.~Abe}
\author[ICRR,IPMU]{K.~Hiraide}
\author[ICRR,IPMU]{K.~Ichimura}
\author[ICRR,IPMU]{Y.~Kishimoto}
\author[ICRR,IPMU]{K.~Kobayashi}
\author[ICRR,IPMU]{M.~Kobayashi}
\author[ICRR,IPMU]{S.~Moriyama}
\author[ICRR,IPMU]{M.~Nakahata}
\author[ICRR]{T.~Norita}
\author[ICRR,IPMU]{H.~Ogawa}
\author[ICRR,IPMU]{H.~Sekiya}
\author[ICRR]{O.~Takachio}
\author[ICRR,IPMU]{A.~Takeda}
\author[ICRR,IPMU]{M.~Yamashita}
\author[ICRR,IPMU]{B.~S.~Yang}

\author[IBS]{N.~Y.~Kim}
\author[IBS]{Y.~D.~Kim}

\author[Gifu]{S.~Tasaka}
\author[Toku]{K.~Fushimi}
\author[IPMU]{J.~Liu}
\author[IPMU]{K.~Martens}
\author[IPMU]{Y.~Suzuki} 
\author[IPMU]{B.~D.~Xu}

\author[Kobe]{R.~Fujita}
\author[Kobe]{K.~Hosokawa}
\author[Kobe]{K.~Miuchi}
\author[Kobe]{Y.~Onishi}
\author[Kobe]{N.~Oka}
\author[Kobe, IPMU]{Y.~Takeuchi}

\author[KRISS, IBS]{Y.~H.~Kim} 
\author[KRISS]{J.~S.~Lee} 
\author[KRISS]{K.~B.~Lee} 
\author[KRISS]{M.~K.~Lee} 

\author[Miyagi]{Y.~Fukuda} 
\author[STE,KMI]{Y.~Itow} 
\author[STE]{R.~Kegasa} 
\author[STE]{K.~Kobayashi}
\author[STE]{K.~Masuda} 
\author[STE]{H.~Takiya} 

\author[Tokai]{K.~Nishijima}

\author[YNU]{S.~Nakamura} 
\address{\rm\normalsize XMASS Collaboration$^*$}

\address[ICRR]{Kamioka Observatory, Institute for Cosmic Ray Research, the University of Tokyo, Higashi-Mozumi, Kamioka, Hida, Gifu, 506-1205, Japan}
\address[IBS]{Center of Underground Physics, Institute for Basic Science, 70 Yuseong-daero 1689-gil, Yuseong-gu, Daejeon, 305-811, South Korea}
\address[Gifu]{Information and multimedia center, Gifu University, Gifu 501-1193, Japan}
\address[Toku]{Institiute of Socio-Arts and Sciences, The University of Tokushima, 1-1 Minamijosanjimacho Tokushima city, Tokushima, 770-8502, Japan}

\address[IPMU]{Kavli Institute for the Physics and Mathematics of the Universe (WPI), the University of Tokyo, Kashiwa, Chiba, 277-8582, Japan}
\address[KMI]{Kobayashi-Maskawa Institute for the Origin of Particles and the Universe, Nagoya University, Furo-cho, Chikusa-ku, Nagoya, Aichi, 464-8602, Japan}

\address[Kobe]{Department of Physics, Kobe University, Kobe, Hyogo 657-8501, Japan}
\address[KRISS]{Korea Research Institute of Standards and Science, Daejeon 305-340, South Korea}
\address[Miyagi]{Department of Physics, Miyagi University of Education, Sendai, Miyagi 980-0845, Japan}
\address[STE]{Solar Terrestrial Environment Laboratory, Nagoya University, Nagoya, Aichi 464-8602, Japan}

\address[Tokai]{Department of Physics, Tokai University, Hiratsuka, Kanagawa 259-1292, Japan}
\address[YNU]{Department of Physics, Faculty of Engineering, Yokohama National University, Yokohama, Kanagawa 240-8501, Japan}

% \altaffiliation{Present address: Kamioka Observatory, Institute for Cosmic Ray Research, the University of Tokyo, Higashi-Mozumi, Kamioka, Hida, Gifu, 506-1205, Japan.}
   
 %\altaffiliation{Present Address: Department of Physics, the University of South Dakota,  Vermillion, SD 57069, USA.}
\cortext[cor1]{{\it E-mail address:} xmass.publications2@km.icrr.u-tokyo.ac.jp .}
\fntext[TASAKA]{Now at Kamioka Observatory, Institute for Cosmic Ray Research, the University of Tokyo, Higashi-Mozumi, Kamioka, Hida, Gifu, 506-1205, Japan.}
\fntext[LIU]{Now at Department of Physics, the University of South Dakota, Vermillion, SD 57069, USA.}

\begin{abstract}
 A search for dark matter was conducted by looking for an annual modulation signal due to the Earth's rotation around the Sun using XMASS, a single phase liquid xenon detector. The data used for this analysis was 359.2 live days times 832 kg of exposure accumulated between November 2013 and March 2015.
When we assume Weakly Interacting Massive Particle (WIMP) dark matter elastically scattering on the target nuclei, 
the exclusion upper limit of the WIMP-nucleon cross section 4.3$\times$10$^{-41}$cm$^2$ at 8 GeV/c$^2$ was obtained and we exclude almost all the DAMA/LIBRA allowed region in the 6 to 16 GeV/c$^2$ range at $\sim$10$^{-40}$cm$^2$. The result of a simple modulation analysis, without assuming any specific dark matter model but including electron/$\gamma$ events, showed a slight negative amplitude. The $p$-values  obtained with two independent analyses are 0.014 and 0.068 for null hypothesis, respectively. we  obtained 90\% C.L. upper bounds that can be used to test various models.
This is the first extensive annual modulation search probing this region with an exposure comparable to DAMA/LIBRA.
\end{abstract}

\begin{keyword}
Dark Matter\sep Annual modulation \sep Liquid xenon


\end{keyword}

\end{frontmatter}


\section{Introduction}
There is  strong evidence that about 5 times more dark matter exists in the universe than ordinary matter. Despite its prominence, we do not yet know what dark matter is \cite{PDG}. Among many candidates for dark matter particles, WIMPs are well motivated and have received the most attention to date.  However, collider experiments at the LHC do not show any indication for such particles so far \cite{PDG}. And no experimental indication for a standard WIMP was found in high sensitivity  direct search experiments such as LUX \cite{LUX}, XENON100 \cite{XENON100} and SuperCDMS \cite{SCDMS} either. On the other hand, that appears to contradict experiments that report signals interpreted as $\sim$ 10 GeV/c$^2$ light WIMP dark matter \cite{SAVAGE,CoGeNT, CDMS-Si} for many years.
 In this situation, light mass WIMPs or other dark matter candidates are getting more attention. In fact, XMASS, a high light yield and low background detector, probed this possibility and looked for signals not only from nuclear recoils but also from electrons and gamma rays emanating from interactions of other candidates such as axion-like particles, Super-WIMPs and so on \cite{XMASS1, XMASS_SW, XMASS4}.

 The most significant result is that of the DAMA/LIBRA experiment at the Gran Sasso National Laboratory in Italy which indicated an annual modulation signature  \cite{dama}.  The Earth's velocity relative to the dark matter distribution changes as the Earth moves around the Sun and produces such a modulation in the dark matter signal rate. This modulation can be observed with terrestrial detectors \cite{Drukier}. 
 The amplitude of the modulation can be changed from positive (i.e. higher rate in June than in December) to negative at cross-over energy \cite{Lewin} and it is possible to observe this effect if the detector threshold is lower than that energy. For 100 GeV/c$^2$ WIMP mass and a Xe target, this is about 20 keV nuclear recoil energy  and it depends on the WIMP mass and the target materials.

The DAMA/LIBRA experiment reported an observation of event rate annual modulation with a 9$\sigma$ significance in 1.33 ton$\cdot$year of data taken over 14 annual cycles with 100 to 250 kg of NaI(Tl) detectors. Their signal may be caused by light WIMPs, or other types of dark matter producing  electrons or gamma rays. In such cases,  the signal is not observable to direct search experiments if they remove electron events. In this situation, dark matter models, for instance, with interaction via dark matter-electron scattering become well motivated which produce keV energy deposition in the detector because they provide a explanation for the DAMA/LIBRA result while avoiding other direct detection constraints \cite{KOPP, FELDSTEIN,FOOT}.
 Recently, in addition to the WIMP search result \cite{XENON100}, an annual modulation search was carried out by the XENON group using only electronic recoil events in their two phase Xe detector with the 34 kg fiducial volume in  224.6 live days data \cite{XENON_MOD}. The result disfavored the interpretation of the DAMA/LIBRA as WIMP-electron scattering through axial-vector coupling.
 XMASS  uses a single phase technology to observe only scintillation light by looking for both types of signals without any electric field. 
Although XMASS has a modest background rate like that of DAMA/LIBRA, XMASS has a larger mass of 832 kg of liquid xenon and,
therefore, is able to reach the DAMA/LIBRA exposure in short time.
While the background in this recent modulation study by the XENON experiment is lower, XMASS has a larger target mass and significantly longer exposure time. We will discuss the sensitivity later.
Note that XMASS tests this modulation hypothesis with almost half the energy threshold ($\sim$ 1keV) than theirs in a different environment and underground site.

\section{The XMASS experiment}

 The XMASS detector is located at the Kamioka Observatory (overburden 2700 m.w.e) in Japan. The detailed design and performance are described in \cite{XMASS_Det}. The detector is immersed in a water tank, 10 m in diameter and 10.5 m in height, which is equipped with 72 Hamamatsu H3600 photomultiplier tubes (PMTs), and acts as an active muon veto and a passive radiation shield against neutrons and gamma rays from the surrounding rock.  642 high quantum efficiency (28-40\% at 175 nm) Hamamatsu R10789 PMTs are mounted in the liquid xenon detector, an approximate sphere with an average radius of 40 cm. The  gain of the PMTs was monitored weekly with a blue LED embedded in the inner surface of the detector. The scintillation light yield response was traced by inserting a $^{57}$Co source \cite{XMASS_Cal} into the detector  every one or two weeks. The number of events for each source position was about 20,000.
%\subsubsection{Data taking and event selection}

 In November 2013, after refurbishing the detector to reduce the radioactive background from the aluminum seal of the PMTs'  window that was identified in the commissioning run \cite{XMASS_Det}, data taking was resumed with about one order of magnitude improved background by covering these seal parts with  plates made of pure copper.  
  The data accumulated between November 2013 and March 2015 were used for this analysis  and we selected periods with stable temperature (172.6-173.0 K) and pressure of Xe (0.162 - 0.164 MPa absolute).  After removing periods of operation with excessive PMT noise or data acquisition problems, the total live time became 359.2 days.

 In this paper,  two different energy scales were used: 1) keV$_{\rm ee}$ represents an electron equivalent energy incorporating all the gamma-ray calibrations in the energy range between 5.9 keV and 122 keV from  $^{55}$Fe, $^{109}$Cd,  $^{241}$Am and $^{57}$Co  sources by inserting sources into the sensitive volume of the detector.  The non-linearity of energy scale was taken into account with those calibrations using a non-linearity model from Doke et al. \cite{DOKE}. Below 5.9 keV, we extrapolated based on this model.  We found about 15\% energy scale difference  from  the Noble Element Simulation Technique (NEST) \cite{NEST} at the threshold energy of 1.1 keV$_{\rm ee}$ ($\sim$8 photoelectrons) in this analysis.
  2) keV$_{\rm nr}$ denotes the nuclear recoil energy which is estimated from the light yield at 122 keV by using non-linearity response measurement at zero electric field in \cite{Leff}. The energy threshold, in this case,  corresponds to 4.8 keV$_{\rm nr}$. 

\section{Data Analysis}
   Events with 4 or more PMT hits in a 200~ns coincidence timing window without a muon veto were initially selected. 
   This  resulted in 3.3$\times10^{7}$ events in the energy region between 1.1 and 15 keV$_{\rm{ee}}$. In order to avoid events caused by afterpulses of bright events induced by, for example, high energy gamma-rays or alpha particles, we rejected  events occurring within 10 ms from the previous event  and having a variance in their hit timings of greater than 100~ns (this selection reduces the number of  events to 2.8$\times10^{7}$).  
   A `Cherenkov cut'  removed events which produce light predominantly from Cherenkov emission, in particular from the beta decays of $^{40}$K in the PMT photocathode. Events for which more than 60\% of their PMT hits arrive in the first 20~ns were classified as Cherenkov-like events  \cite{XMASS1} (this selection reduces the number of events to 1.9$\times10^{6}$).  Finally, to remove background events that occurred in front of PMT window, we give upper limits on the values of `Max-photoelectron/Total-photoelectron'  where Max-photoelectron and Total-photoelectron are the largest photoelectron counts in one PMT among all PMTs and the total number of photoelectrons in the event, respectively (this selection reduces the number of events to 3.6$\times10^{5}$). These cut values varied as a function of photoelectron from about 0.2 at 8 photoelectrons to about 0.07 at 50 photoelectrons. The count rate for the data after all the cuts is 1.17 (0.028) events/day/kg/keV$_{\rm ee}$ at 1.1 (5.0) keV$_{\rm ee}$.
\begin{figure}[htbp]
\centering
\includegraphics[width=0.5\textwidth]{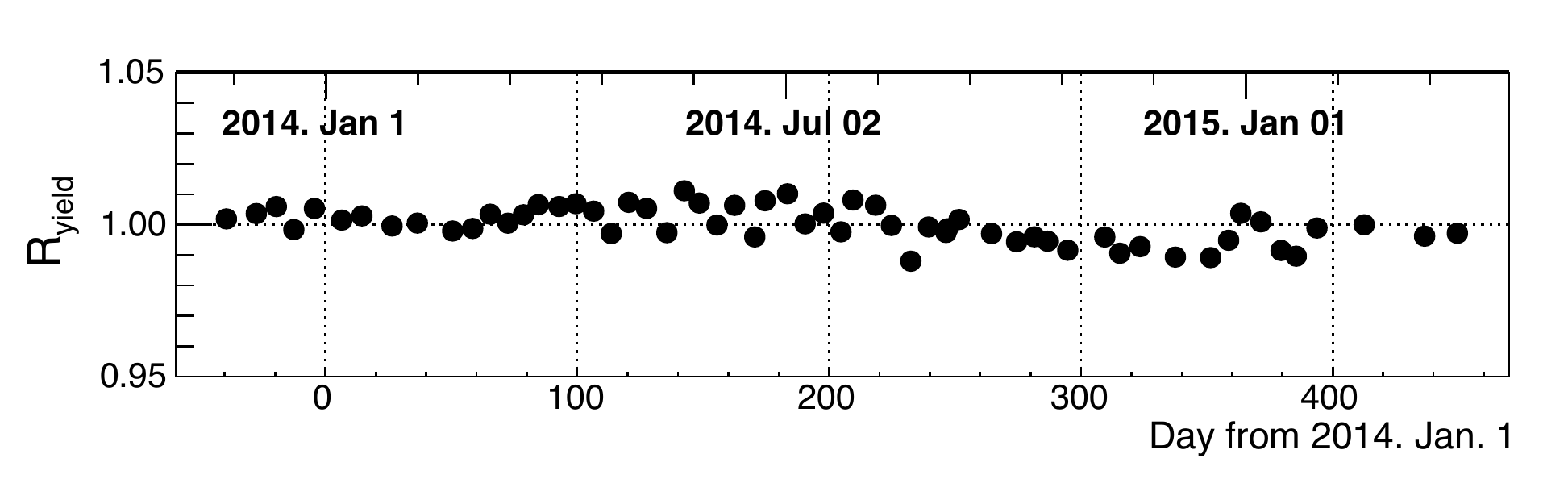}
\caption{Light yield stability was  monitored with a $^{57}$Co 122 keV gamma ray source. The relative intrinsic scintillation light yield ($R_{yield}$) was obtained by comparing to calibration data with the Monte Carlo simulation by considering optical parameters such as absorption and scattering length.}
\label{fig:stability}
\end{figure}

The $^{57}$Co calibration data were taken at from $z=-$40 cm to $+$40 cm along the center vertical axis of the detector to track photoelectron yield and optical properties of the liquid xenon \cite{XMASS_Det}. A difference of about 10\% was observed as the position dependence for this photoelectron yield. The photoelectron yield during the data taking varied about 10\%.
The absorption and scattering length for the scintillation light as well as the intrinsic light yield of the liquid xenon scintillator are extracted from the $^{57}$Co calibration data the Monte Carlo simulation \cite{XMASS_Det}. 
With that we found that we can trace the observed photoelectron change in the calibration data as a change as the absorption length, while the scattering length remains stable at 52 cm with a standard deviation of  $\pm 0.6\%$. We then re-evaluate the absorption length and the relative intrinsic light yield to see the stability of the scintillation light response by fixing the scattering length at 52 cm. The absolute absorption length varied from about 4 m to 11 m, but the relative intrinsic light yield ($R_{yield}$) stayed within $\pm 0.6\%$ over the entire data taking period.

\begin{figure}[htbp]
\centering
\includegraphics[width=0.5\textwidth]{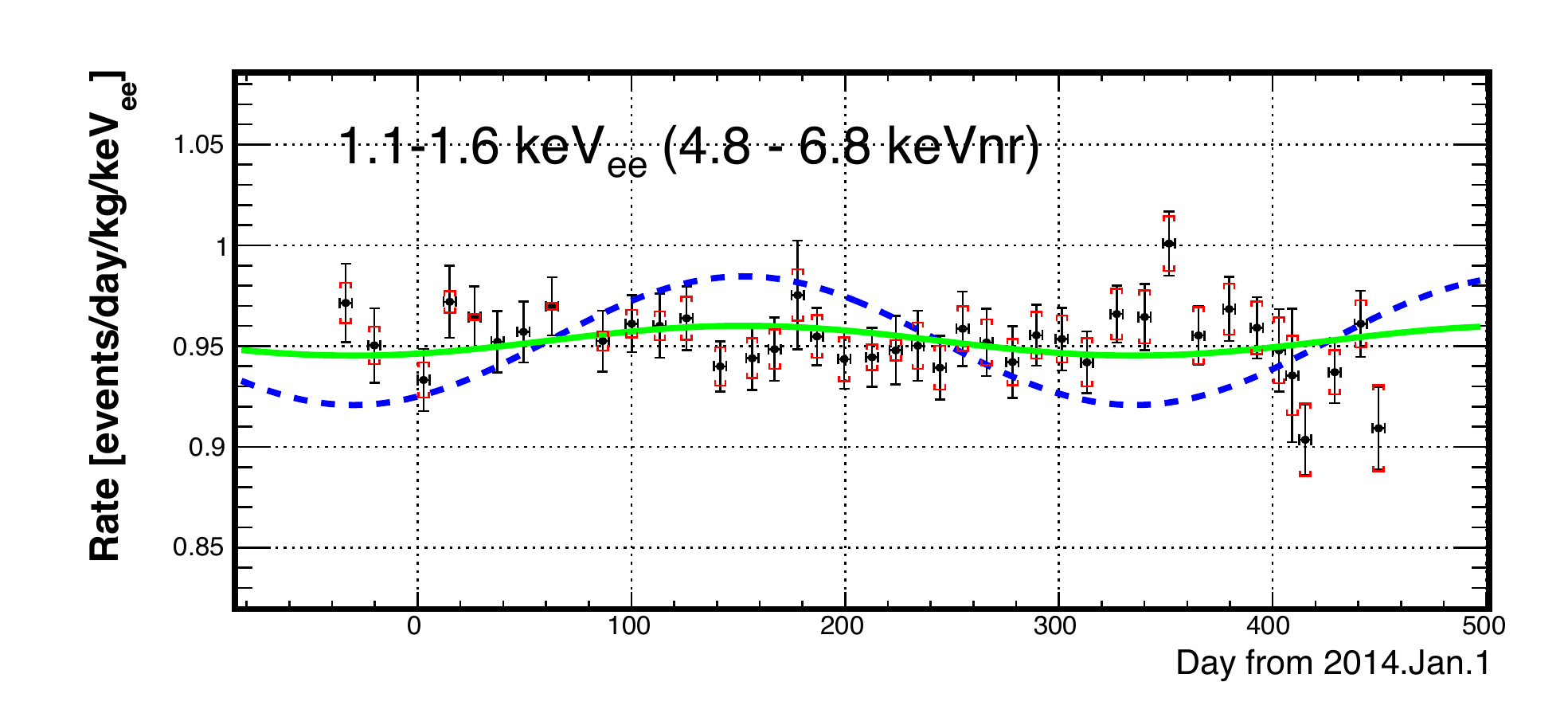}
\caption{(Color online) Observed count rate as a function of time in the 1.1 - 1.6 keV$_{\rm ee}$ (= 4.8 - 6.8 keV$_{\rm nr}$) energy range. The black error bars show the statistical uncertainty of the count rate. Square brackets indicate the 1$\sigma$ systematic error for each time bin.
The solid and dashed  curves indicate the expected count rates assuming 7 and 8 GeV/c$^{2}$ WIMPs  respectively with a cross section of 2$\times10^{-40} \rm{cm}^{2}$ where the WIMP search sensitivity closed to DAMA/LIBRA.}
\label{fig:rate}
\end{figure}

 The time dependence of the photoelectron yield affects the efficiency of the cuts. Therefore, we evaluate the absorption length dependence of the relative cut efficiencies through Monte Carlo simulation. If we normalize the overall efficiency at an absorption length of 8 m, this efficiency changes from $-$4\% to +2\% over the relevant absorption range. The position dependence of the efficiency was taken into account as a correlated systematic error ($\sim \pm 2.5\%$). This is the dominant systematic uncertainty in the present analysis. The second largest contribution comes from a gain instability of the waveform digitizers (CAEN V1751)  between April 2014 and September 2014 due to a different calibration method of the digitizers used in that period. This effect  contributes an uncertainty of 0.3\% to the energy scale. Other effects from LED calibration, trigger threshold stability, timing calibration were negligible. The observed count rate after cuts as a function of time in the energy region between 1.1 and 1.6 keV$_{\rm{ee}}$ is shown in Fig.~\ref{fig:rate}. The systematic errors caused by the relative cut efficiencies are also shown.

 To retrieve the annual modulation amplitude from the data, the least squares method for the time-binned data was used. The data set was divided into 40 time-bins ($t_{bins}$) with roughly 10 days of live time each.
The data in each time-bin were then further divided into energy-bins  ($E_{bins}$) with a width of 0.5 keV$_{\rm ee}$. Two fitting methods were performed independently.  Both of them fit all energy- and time-bins simultaneously.
 Method 1 used a `pull term' $\alpha$ with $\chi^{2}$ defined as:
\begin{equation}
\chi^2 = \sum\limits_{i}\limits^{E_{bins}} \sum\limits_{j}\limits^{t_{bins}} 
\left(\frac{(R^{{\rm data}}_{i,j}-R^{\rm ex}_{i,j} - \alpha K_{i,j})^2}{\sigma({\rm stat})^2_{i,j}+\sigma({\rm sys})^2_{i,j}}\right)+\alpha^{2}, 
\end{equation}
where $R_{i,j}^{\rm data}$, $R_{i,j}^{\rm ex}$ , $\sigma(\rm{stat)}_{i,j}$ and $\sigma(\rm{sys)}_{i,j}$ are data, expected
event rate, statistical and systematic error, respectively, of the ($i$-th energy- and $j$-th time-) bin. The time is denoted as the number of days from January 1, 2014. $K_{i,j}$ represents the 1$\sigma$ correlated systematic error on the expected event rate based on the relative cut efficiency in that bin. Method 2 used a covariance matrix to propagate the effects of the systematic error. Its $\chi^{2}$ was defined as:
\begin{equation}
\chi^2 =\sum_{k,l}^{N_{\rm bins}} (R_{k}^{\rm data}-R_{k}^{\rm ex}) 
(V_{\rm stat}+V_{\rm sys})^{-1}_{kl} 
(R_{l}^{\rm data}-R_{l}^{\rm ex}), \\ 
\end{equation}
where $N_{\rm bins} (= E_{bins} \times t_{bins})$ was the total number of  bins and
$R_{k}^{\rm data(ex)}$ is the event rate where $k = i\cdot t_{bins}+j$.
 The matrix $V_{\rm stat}$  contains the statistical uncertainties of the bins, and $V_{\rm sys}$ is the covariance matrix of the systematic uncertainties as derived from the relative cut efficiency. 

\section{Results and Discussion}
 We performed two analyses, one assuming WIMP interactions and the other  independent of any specific dark matter model. Hereafter we call the former case the WIMP analysis and the latter  a model independent analysis. 
 
\begin{figure}[htbp]
\centering
\includegraphics[width=0.5\textwidth]{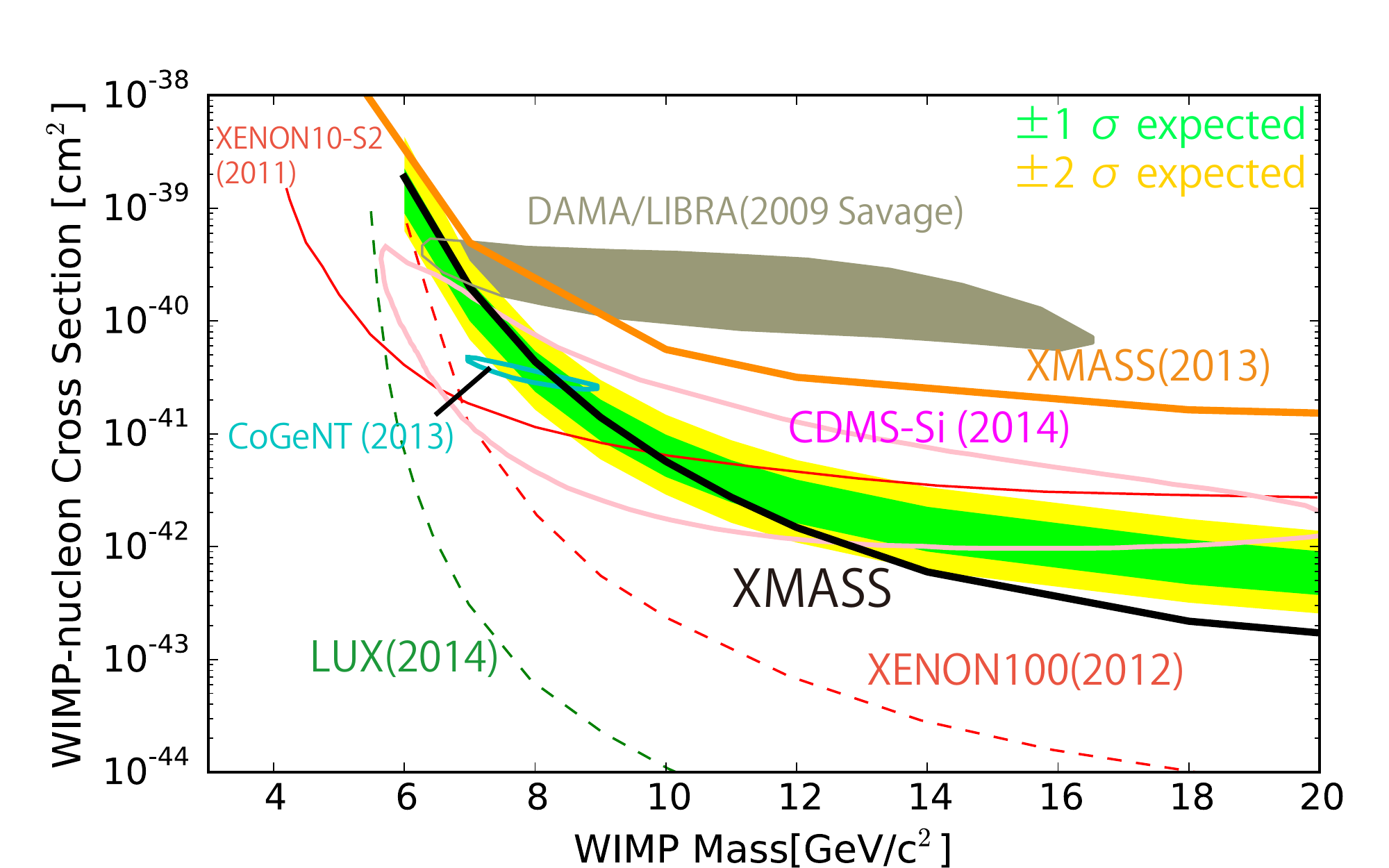}
\caption{(Color online)  Limits on the spin-independent elastic WIMP-nucleon cross section as a function of WIMP mass. The solid line shows the XMASS 90\% C.L. exclusion from the  annual modulation analysis. The $\pm1\rm{\sigma}$ and $\pm2\rm{\sigma}$ bands represent the expected 90\% exclusion distributions. Limits  as well as allowed regions from other searches based on counting method are also shown \cite{LUX, XENON100,Xe10, SAVAGE, CoGeNT,CDMS-Si,XMASS1}. }
\label{fig:MD}
\end{figure}

  In the case of the WIMP analysis, the expected modulation amplitudes  become a function of the WIMP mass $A_i(m_\chi)$ as the WIMP mass $m_\chi$ determines the recoil energy spectrum.  The expected rate in a bin then becomes:
\begin{equation}
R_{i,j}^{\rm ex}= \int_{t_{j}-\frac{1}{2}\Delta t_{j}}^{t_{j}+\frac{1}{2} \Delta t_{j}} \bigl(C_{i} + \sigma_{\chi n} \cdot A_{i}(m_{\chi}) \cos 2\pi \small{\frac{(t-t_{0})}{T}} \bigr) dt,
\label{eq:MD}
\end{equation}
where  $\sigma_{\chi n}$ is  the WIMP-nucleon cross section. To obtain the WIMP-nucleon cross section the data was fitted in the energy range of  1.1-15 keV$_{\rm{ee}}$.  We assume a standard spherical isothermal galactic halo model with the most probable speed of $v_{0}$=220~km/s, the Earth's 
velocity relative to the dark matter distribution of $v_{E} = 232 + 15 ~{\rm sin} 2\pi(t - t_{0})/T$~km/s, and a galactic escape velocity of $v_{esc}$ = 650 km/s, a local dark matter density of 0.3 GeV/cm$^{3}$, following \cite{Lewin}. 
In the analysis, the signal efficiencies for each WIMP mass are estimated from Monte Carlo simulation of uniformly distributed nuclear recoil events in the liquid xenon volume.
The systematic error of the efficiencies comes from the uncertainty of liquid xenon scintillation decay time of 25$\pm$1~ns \cite{XMASS1} and  is estimated as about 5\% in this analysis.
The expected count rate for WIMP masses of 7 and 8 GeV/c$^{2}$ with a cross section of 2$\times10^{-40}$ cm$^{2}$  for the spin independent case  are shown in Fig.~\ref{fig:rate} as a function of time after all cuts. This demonstrates the high sensitivity of the XMASS detector to modulation.
As both methods found no significant signal, the 90\% C.L. upper limit by  the `pull term' method on the WIMP-nucleon cross section is shown in Fig.~\ref{fig:MD}. The exclusion upper limit of 4.3$\times10^{-41} \rm{cm}^{2} $ at 8 GeV/c$^{2}$ was obtained.  The $-1\sigma$ scintillation efficiency of \cite{Leff} was used  to obtain a conservative limit.
To evaluate the sensitivity of WIMP-nucleon cross section, we carried out a statistical test by applying the same analysis to 10,000 dummy samples with the same statistical and systematic errors as data but without modulation by the following  procedure.
 At first, the time-averaged energy spectrum was obtained from the observed data. Then, we performed a toy Monte Carlo simulation to simulate time variation of event rate of background at each energy bin assuming the same live time as data and including systematic uncertainties.
 The $\pm1 \sigma$ and  $ \pm2\sigma$ bands in Fig.~\ref{fig:MD} outline the expected 90\% C.L. upper limit band for the no-modulation hypothesis using the dummy samples.
The result excludes the DAMA/LIBRA allowed region as interpreted  in  \cite{SAVAGE} for the WIMP masses higher than 8 GeV/c$^{2}$. The difference  between two fitting methods is less than 10\%.   The upper limit of 5.4$\times10^{-41} \rm{cm}^{2}$  is obtained  under different astrophysical assumptions  of  $v_{esc}$ = 544~km/s \cite{RAVE}.
 The best fit parameters in a  mass range between 6 and 1000 GeV/c$^2$ is a cross section of 3.2$\times10^{-42}$ $\rm{cm}^{2}$ for a WIMP mass of 140 GeV/c$^{2}$. This yields a  statistical significance of 2.7$\sigma$, however, in this case, the expected unmodulated event rate  exceeds the total observed event rate by a factor of 2, therefore these parameters were deemed unphysical.   

\begin{figure}[htbp]
\centering
\includegraphics[width=0.5\textwidth]{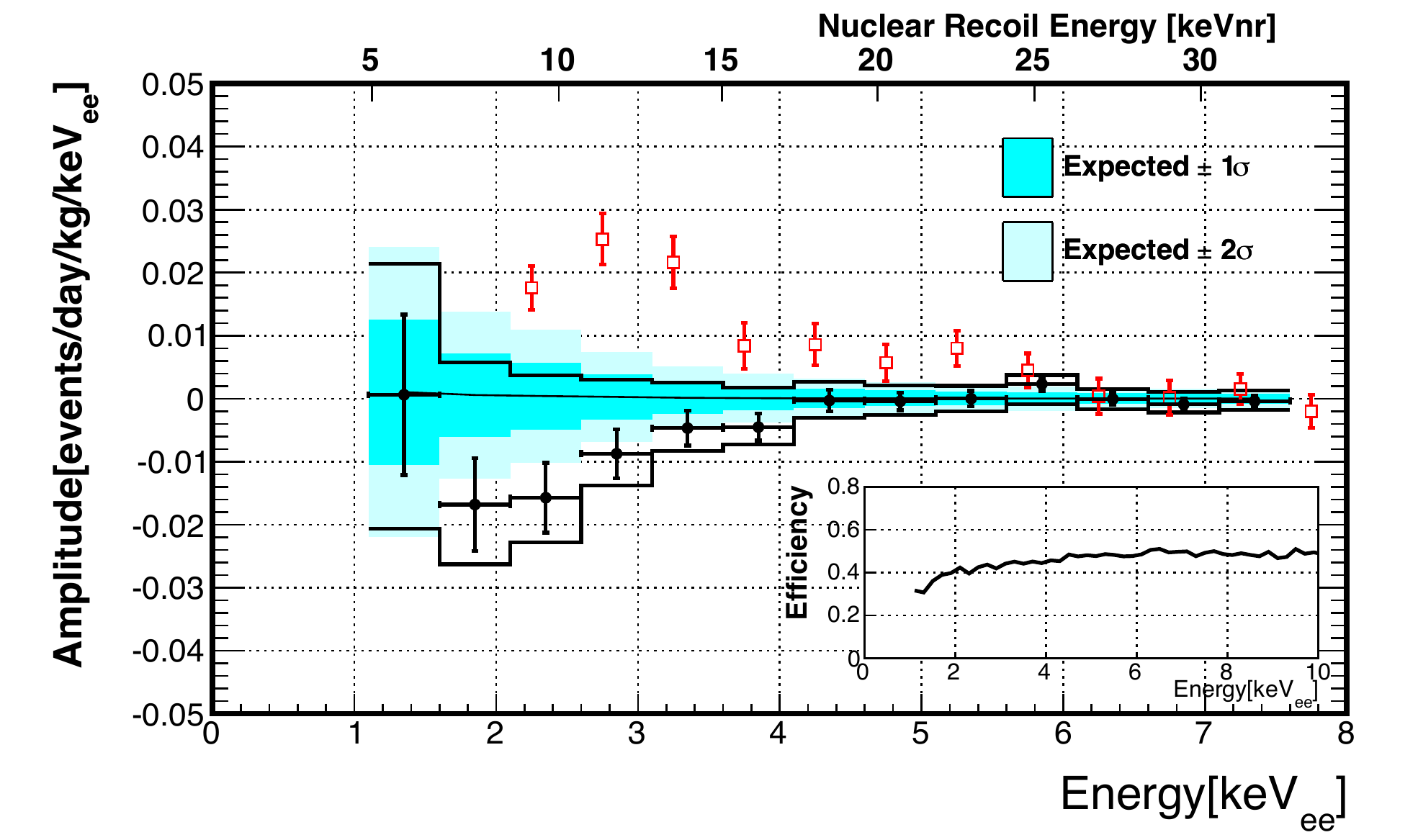}
\caption{ (Color online) Modulation amplitude  as a function of energy for the model independent analyses using  the `pull term' method (solid circle). Solid lines represent  90\% positive (negative) upper limits on the amplitude.  The $\pm1\rm{\sigma}$ and $\pm2\rm{\sigma}$ bands represent the expected amplitude region (see detail in the text). DAMA/LIBRA result (square) is also shown \cite{dama}. }
\label{fig:MI}
\end{figure}

 For the model independent analysis, the expected event rate was estimated as:
\begin{equation}
R_{i,j}^{\rm ex} = \int_{t_{j}-\frac{1}{2}\Delta t_{j}}^{t_{j}+\frac{1}{2} \Delta t_{j}} \left( C_{i} +A_{i} \cos 2\pi \small{\frac{(t-t_{0})}{T}} \right) dt, 
\label{eq:MI}
\end{equation}
where the free parameters $C_{i}$ and $A_{i}$ were the unmodulated event rate and the modulation amplitude, respectively.  $t_{0}$ and $T$ were the phase and period of the modulation, and $t_{j}$ and $\Delta t_{j}$ was the time-bin's center and width, respectively.  In the fitting procedure,  the 1.1--7.6 keV$_{\rm ee}$ energy range was used and the modulation period $T$ was fixed to one year and the phase $t_{0}$ to 152.5 days ($\sim$2nd of June)  when the Earth's velocity relative to the dark matter distribution is expected to be maximal.
 Figure~\ref{fig:MI} shows the best fit amplitudes as a function of energy for `pull term' after correcting the efficiency. The efficiency was evaluated from gamma ray Monte Carlo simulation with a flat energy spectrum uniformly distributed in the sensitive volume (Fig.~\ref{fig:MI} inset).   Both methods are in good agreement and find a slight negative amplitude below 4~keV$_{\rm{ee}}$. 
  The $\pm1\rm{\sigma}$ and $\pm2\rm{\sigma}$ bands in Fig.~\ref{fig:MI} represent expected amplitude coverage derived from same dummy sample above by the `pull term' method. This test gave a $p$-value of 0.014 (2.5$\sigma$)  for the `pull term' method and of 0.068 (1.8$\sigma$) for the covariance matrix method.  
To be able to test any model of dark matter, we evaluated the constraints on the positive and negative amplitude separately in Fig.~\ref{fig:MI}.
 The upper limits on the amplitudes in each energy bin were calculated by considering only regions of positive or negative amplitude. They were calculated by  integrating Gaussian distributions based on the mean and sigma of data (=$G(a)$) from zero.  The positive or  negative upper limits are satisfied with 0.9 for  $\int_0^{a_{up}} G(a)da/\int_0^\infty G(a)da$ or $\int_{a_{up}}^{0} G(a)da/\int_{-\infty}^0 G(a)da$, where $a$ and $a_{up}$  are the amplitude and its 90\% C.L. upper limit, respectively.
 The `pull term' method obtained   positive (negative) upper limit of $2.1 (-2.1)\times10^{-2}$ events/day/kg/keV$_{\rm{ee}}$  between 1.1 and 1.6 keV$_{\rm{ee}}$ and the limits become stricter at  higher energy.  The energy resolution ($\sigma/$E) at 1.0 (5.0) keV$_{\rm{ee}}$ is estimated to be 36\% (19\%) comparing gamma ray calibrations and its Monte Carlo simulation.
 As a guideline, we make direct comparisons with other experiments not by considering a specific dark matter model but amplitude count rate. 
The modulation amplitude of $\sim2 \times10^{-2}$ events/day/kg/keV$_{\rm ee}$ between 2.0 and 3.5 keV$_{\rm ee}$ was obtained by DAMA/LIBRA \cite{dama} and we estimate  a 90\% C.L. upper limit for XENON100 as $3.7\times10^{-3}$ events/day/kg/keV$_{\rm ee}$ (2.0--5.8  keV$_{\rm ee}$) based on \cite{XENON_MOD} as it was not claimed as a signal. XMASS obtained positive upper limits of $(1.7-3.7)\times10^{-3}$ events/day/kg/keV$_{\rm ee}$ in same energy region and gives the more stringent constraint. This fact is important when we test the dark matter model. 

\section{Conclusions}
 In conclusion, XMASS with its large exposure and high photoelectron yield (low energy threshold) conducted an annual modulation search. For the WIMP analysis,  the exclusion upper limit of 4.3$\times10^{-41} \rm{cm}^{2} $ at 8 GeV/c$^{2}$ was obtained and the result excludes the DAMA/LIBRA allowed region for WIMP masses higher than that.
 In the case of the model independent case,  the analysis was carried out from the energy threshold of 1.1 keV$_{\rm ee}$  which is lower than DAMA/LIBRA and XENON100. The positive (negative) upper limit amplitude of $2.1~(-2.1) \times10^{-2}$ events/day/kg/keV$_{\rm{ee}}$  between 1.1 and 1.6 keV$_{\rm{ee}}$ and $(1.7-3.7)\times$10$^{-3}$ counts/day/kg/keV$_{\rm ee}$ between 2 and 6 keV$_{\rm ee}$ were obtained.
As this analysis does not consider only nuclear recoils, a simple electron or gamma ray interpretation of  the DAMA/LIBRA signal can also obey this limit.

\section*{Acknowledgments}
We gratefully acknowledge the cooperation of Kamioka Mining
and Smelting Company. 
This work was supported by the Japanese Ministry of Education,
Culture, Sports, Science and Technology, Grant-in-Aid
for Scientific Research, 
JSPS KAKENHI Grant Number, 19GS0204, 26104004, and partially
by the National Research Foundation of Korea Grant funded
by the Korean Government (NRF-2011-220-C00006).





\end{document}